\documentclass{article}

\usepackage{amssymb}
\usepackage{amsmath}
\usepackage{epsfig}
\usepackage{meta}

\title{Wide-Band Tuneability, Nonlinear Transmission, and Dynamic Multistability in SQUID Metamaterials} 

\makeatletter
\def\name#1{\gdef\@name{#1\\}}
\makeatother
\name{{\bf \large G. P. Tsironis, N. Lazarides, I. Margaris}} 

\address{
Crete Center for Quantum Complexity and Nanotechnology, Department of Physics, \\
University of Crete, P.O. Box 2208, 71003 Heraklion, Greece \\
$\&$ \\
Institute of Electronic Structure and Laser, \\
Foundation for Research and Technology-Hellas, P.O. Box 1527, 71110 Heraklion, Greece \\
*corresponding author, E-mail: {\tt nl@physics.uoc.gr}
}
\begin{document}
\maketitle
\begin{abstract}
Superconducting metamaterials comprising rf SQUIDs  
(Superconducting QUantum Interference Devices) have been recently realized and investigated 
with respect to their tuneability, permeability and dynamic multistability properties.
These properties are a consequence of intrinsic nonlinearities due to the sensitivity
of the superconducting state to external stimuli. SQUIDs, made of a superconducting 
ring interrupted by a Josephson junction, possess yet another source of nonlinearity,
which makes them widely tuneable with an applied dc dlux. A model SQUID metamaterial, 
based on electric equivalent circuits, is used in the weak coupling approximation
to demonstrate the dc flux tuneability, dynamic multistability, and nonlinear transmission
in SQUID metamaterials comprising non-hysteretic SQUIDs.
The model equations reproduce the experimentally observed tuneability patterns,
and predict tuneability with the power of an applied ac magnetic magnetic field. 
Moreover, the results indicate the opening of nonlinear frequency bands for energy
transmission through SQUID metamaterials, for sufficiently strong ac fields.
\end{abstract}
%
\section{Introduction}
{\em Superconducting metamaterials} \cite{Anlage2011}, a particular class of artificial media
that rely on the extraordinary properties of superconductors at sufficiently low 
temperatures, have been recently attracted great attention (see e.g., reference \cite{SUST-Focus}).
Conventional metamaterials, that comprise highly conducting metallic elements 
\cite{Smith2004,Shalaev2007,Soukoulis2011}, typically exhibit high losses in the
frequency range where their unusual and sought properties are manifested.
The key element for the construction of conventional (metallic) metamaterials has
customarily been the split-ring resonator (SRR), typically a highly conducting metallic 
ring with a slit, that can be regarded as an inductive-capacitive ($L\, C$) resonant 
oscillator. Nonlinearity, provided by combination of SRRs with electronic components
(e.g., diodes \cite{Shadrivov2008}), adds a new degree of freedom for the design of tuneable
metamaterials. Superconductors, on the other hand, are intrinsically nonlinear materials,
due to the extreme sensitivity of the superconducting state in external stimuli 
\cite{Zheludev2010,Zheludev2011}, which moreover exhibit significantly reduced Ohmic losses.
They thus provide unique opportunities to the researchers in the field for the fabrication 
of superconducting metamaterials with highly controllable effective electromagnetic 
properties including wideband tuneability
\cite{Ricci2005,Ricci2007,Chen2010,Jin2010,Fedotov2010,Kurter2012,Singh2012a,Singh2012b,Savinov2012,Jung2013,Butz2013a,Zueco2013}.

The direct superconducting analogue of a nonlinear SRR is the rf SQUID (rf Superconducting
QUantum Interference Device), a long-known device in community of superconductivity.
It consists of a superconducting ring interrupted by a Josephson junction (JJ)
Josephson junction (JJ) \cite{Josephson1962}, as shown schematically in figure 1(a). 
The SQUID is a strongly nonlinear resonator \cite{Barone1982,Likharev1986},
that is tuneable over a wide frequency range by applying either a flux bias 
\cite{Jung2013,Trepanier2013} or by varying the incoming rf power of an applied alternating
field \cite{Trepanier2013,Jung2013b}. This superconducting device has found up to date 
numerous applications \cite{Jenks1997,Koelle1999,Kleiner2004,Fagaly2006}, and it is known 
to be the world’s most sensitive detector of magnetic signals. 
The replacement of metallic SRRs with rf SQUIDs, which have no direct electrical conduct 
but instead they are coupled magnetically through their mutual inductances,
has been suggested theoretically a few years ago \cite{Du2006,Lazarides2007}. 
Such SQUID metamaterials have been recently realized in the lab 
\cite{Jung2013,Butz2013a,Butz2013b,Trepanier2013,Jung2013b},  
that exhibit strong nonlinearities and wide-band tuneability with unusual magnetic properties
due to macroscopic quantum effects.
Nonlinearity and discreteness in SQUID metamaterials, along with low Ohmic losses (at least 
at microwave frequencies) may also lead in the generation of discrete breathers
\cite{Lazarides2008a,Tsironis2009,Lazarides2012},
i.e., time-periodic and spatially localized modes that change locally the magnetic response.
Recent advances that led to nano-SQUIDs makes possible the fabrication of SQUID metamaterials
at the nanoscale \cite{Wernsdorfer2009} (and references therein).

Superconducting metamaterials, resulting either from bare replacement of the metallic parts
employed in conventional metamaterials by superconducting ones
\cite{Gu2010,Jin2010,Chen2010,Zhang2011,Wu2011a,Zhang2012,Zhang2013b,Zhang2013a},
or by constructing hybrid structures comprising superconducting components
\cite{Pimenov2005,Kussow2007,Limberopoulos2009,Rakhmanov2010,Golick2010,Kurter2011,Kurter2012,Prat-Camps2013},
or even by incoprorating the Josephson effect as in arrays of SQUIDs 
\cite{Castellanos2008,Jung2013,Butz2013a,Butz2013b,Trepanier2013}, 
have been already proposed and/or demonstrated experimentally.
Their operation frequency spans a huge range, from zero \cite{Magnus2008,Navau2009,Mawatari2012,Mawatari2013}
to microwaves \cite{Kurter2010,Jung2013,Ghamsari2013,Butz2013a,Butz2013b}
and to Terahertz
\cite{Chen2010,Jin2010,Gu2010,Wu2011b,Zhang2011,Wu2011a,Zhang2012,Singh2012a,Tsiatmas2012,Zuo2012,Zhang2013b,Zhang2013a}
and visible \cite{Limberopoulos2009} frequencies.
Moreover, researchers rely on particular superconducting devices to access the trully
quantum metamaterial regime \cite{Du2006,Zagoskin2012,Mukhin2013,Wilson2013,Macha2013}.

In the present work, numerical calculations that rely on a model SQUID metamaterial in the
weak coupling approximation are shown to reproduce fairly well the experimentally observed
tuneability patterns with an applied constant (dc) flux bias.
Specifically, it is demonstrated that the frequency band of linear flux waves 
of a SQUID metamaterial can be tuned periodically with the dc flux, with a period of one
flux quantum. These results in a sense validate the discrete model.
Quantitative differences in the tuneability patterns are attributed to the system 
size, dimensionality, nonlinearity strength (equivalently the amplitude of an applied rf
alternating magnetic field), and coupling strength between neighboring SQUIDs. 
The transmission properties of SQUID metamaterials are investigated in the weakly nonlinear 
regime for a one-dimensional (1D) SQUID metamaterial with respect to the loss coefficient
of the SQUIDs, for coupling strength which is in accordance with the weak coupling 
approximation (and in the range of values achieved in the experiments).
For low losses and substantial nonlinearity, the transmission of flux waves through the 
metamaterial is also possible for frequencies outside the linear band, though the opening 
of one or more "channels" (i.e., nonlinear bands), whose effectiveness in transmitting 
energy depends on the nonlinearity strength.
\begin{figure}[!h]
\centerline{\epsfig{figure=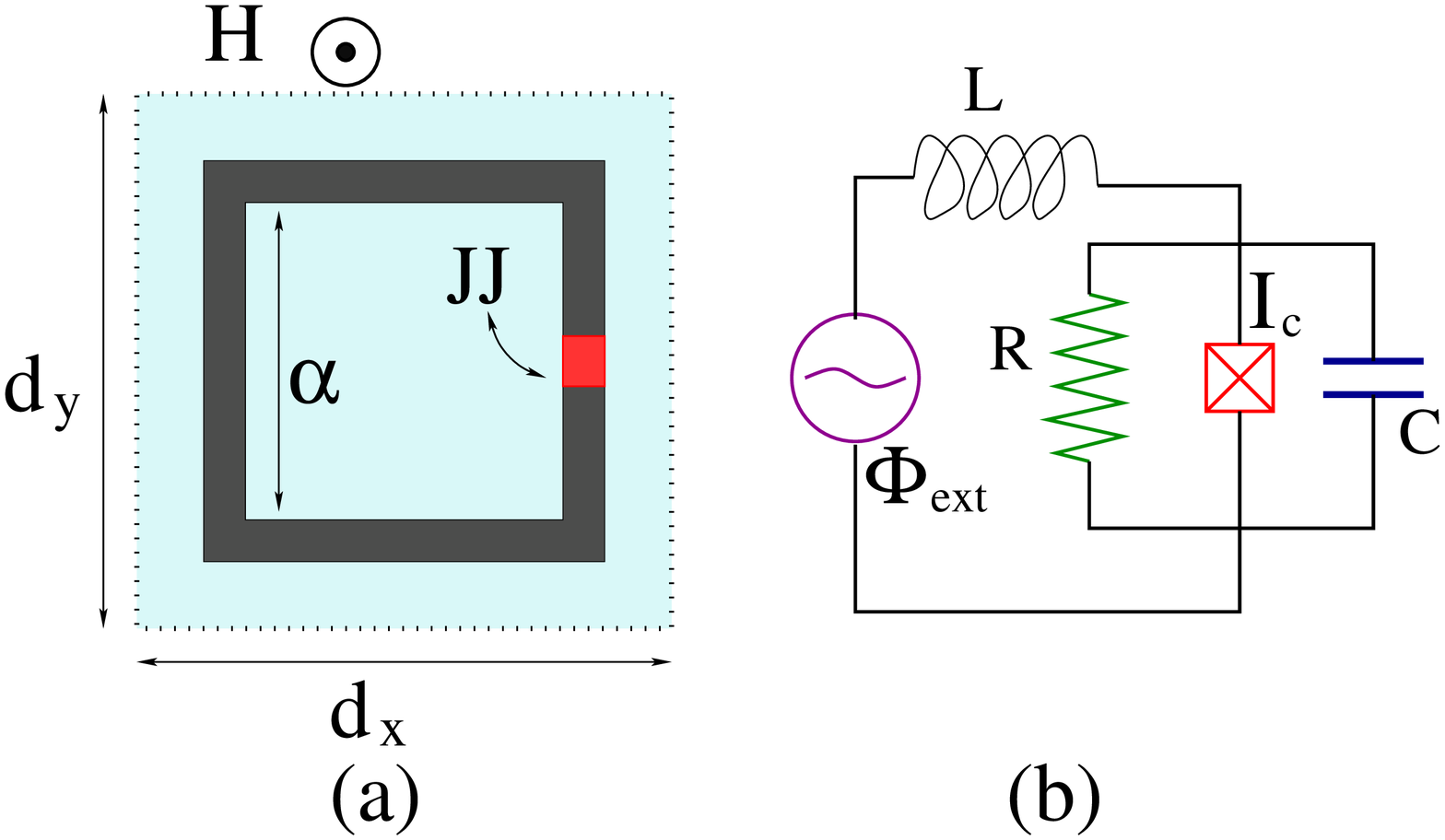,width=75mm}}
\centerline{\epsfig{figure=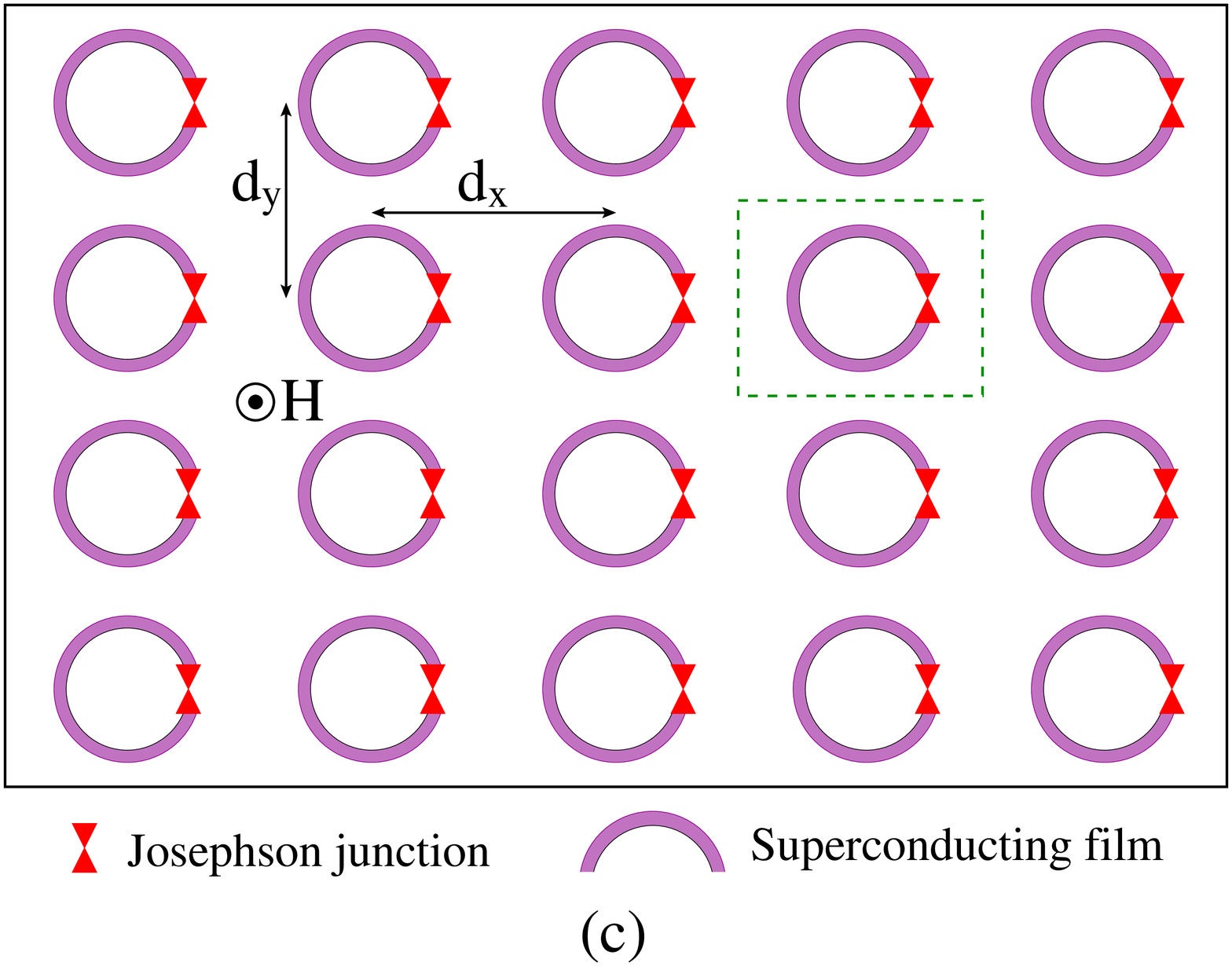,width=75mm}}
\caption{(color online)
(a) Schematic drawing of an rf SQUID in a perpendicular time-dependent magnetic field
${\bf H}(t)$.
(b) Equivalent electrical circuit (RCSJ model) for a single rf SQUID driven by a
flux source $\Phi_{ext}$.
(c) Schematic drawing of a two-dimensional rf SQUID metamaterial, where the unit cell
is indicated by the green-dashed line.
}
\label{schematic-squid-array}
\end{figure}

\section{SQUID Metamaterial Equations}
Consider a two-dimensional (2D) rectangular lattice of identical SQUIDs in an ac and/or
dc magnetic field, ${\bf H} (t)$, which is spatially uniform, as in figure 1(c).
In the following we adopt the description
(and notation) in references \cite{Lazarides2012,Lazarides2013b}, thus summarizing the 
essential building blocks of the model in a self-contained manner, 
yet omitting unnecessary details.
The dynamic equations for the fluxes, $\phi_{n,m}$, in the $(n,m)$ SQUID ring
are given by \cite{Lazarides2008a,Tsironis2009,Lazarides2013b}
\begin{eqnarray}
\label{2.01}
  \ddot{\phi}_{n,m} +\gamma \dot{\phi}_{n,m} +\phi_{n,m} +\beta\, \sin( 2 \pi \phi_{n,m} )
\nonumber \\
  -\lambda_x ( \phi_{n-1,m} +\phi_{n+1,m} ) 
  -\lambda_y ( \phi_{n,m-1} +\phi_{n,m+1} )
\nonumber \\
   = [1-2(\lambda_x +\lambda_y)] \phi_{ext},
\end{eqnarray}
where the overdots denote differentiation with respect to the normalized time $\tau$,
$\lambda_x$ and $\lambda_y$ are the coupling coefficients in the $x$ and $y$ direction,
respectively, $\beta ={L I_c}/{\Phi_0} =\beta_L /2\pi$ is the rescaled SQUID parameter,
and $\gamma$ is a dimensionless coefficient representing all of the dissipation in 
each SQUID.
The frequency and time in equation (\ref{2.01}) are normalized to the corresponding $L C$ 
SQUID frequency $\omega_0 =1/\sqrt{LC}$ and its inverse $\omega_0^{-1}$, respectively,
while all fluxes are normalized to the flux quantum, $\Phi_0$.
The external flux $\phi_{ext}$ is of the form
\begin{eqnarray}
\label{2.02a}
  \phi_{ext} = \phi_{dc}  +\phi_{ac} \cos(\Omega \tau ) , 
\end{eqnarray}
where $\Omega$ is the driving frequency, $\phi_{dc}$ is the flux bias, and $\phi_{ac}$
is the amplitude of the ac field. The current, $i_{n,m}$, flowing in the $(n,m)$ SQUID,
normalized to the critical current of the junction, $I_c$, is given by
\begin{eqnarray}
\label{2.03}
   i_{n,m} =\frac{1}{\beta} \left\{ \phi_{n,m} - \phi_{eff} 
      -\lambda_x ( \phi_{n-1,m} + \phi_{n+1,m} ) 
       \right.
\nonumber \\ \left.
      -\lambda_y ( \phi_{n,m-1} + \phi_{n,m+1} ) \right\}.
\end{eqnarray}
For later use, the total maximum current per SQUID of the SQUID metamaterial is defined as
\begin{eqnarray}
\label{2.031}
   i_{max} = max\left\{ \frac{1}{N_x N_y} \sum_{n,m=1}^{N_x,N_y} i_{n,m} \right\}_T,
\end{eqnarray}
i.e., the maximum value of the total current (given by the expression in the curly brackets)
in one driving period $T=2\pi/\Omega$.  
The SQUID metamaterial supports linear flux waves which frequency dispersion is
\begin{eqnarray}
\label{2.04}
   \Omega \equiv \Omega_{\bf \kappa} = \sqrt{1 + \beta_L -2( \lambda_x \, \cos \kappa_x
                                +\lambda_y \, \cos \kappa_y ) }  , 
\end{eqnarray}
where ${\bf \kappa} =(\kappa_x, \kappa_y)$ is the normalized wavevector.

The dynamic behavior of rf SQUIDs has has been investigated extensively for more than
two decades, both in the hysteretic ($\beta_L >1$) and the non-hysteretic ($\beta_L <1$)
regimes. Just as conventional metamaterials acquire their properties from their individual 
elements, SQUID metamaterials acquire their resonance properties from individual SQUIDs.
For very low amplitude of the ac driving field (linear regime), an rf SQUID exhibits
a resonant magnetic response at a particular frequency
\begin{equation}
\label{2.05}
  \omega_{SQ} = \omega_0 \sqrt{ 1 +\beta_L } ,
\end{equation}
which is always larger than its corresponding $L C$ frequency, $\omega_0$.
Tuneability of an rf SQUID can be achieved either with the amplitude of the ac field, 
$\phi_{ac}$, or with the flux bias, $\phi_{dc}$, which generate the flux threading its loop
\cite{Lazarides2012}. The resonance shift due to a dc flux bias has been actually observed
in low$-T_c$, single rf SQUIDs in the linear regime \cite{Jung2013}, as well as in 1D SQUID
metamaterials \cite{Butz2013a,Butz2013b,Ghamsari2013}.
Assuming isotropic coupling, i.e., $\lambda_x =\lambda_y =\lambda$, 
the maximum and minimum values of the linear frequency band are then obtained by
substituting ${\bf \kappa} =(\kappa_x, \kappa_y) =(0, 0)$ and $(\pi, \pi)$, respectively,
into equation (\ref{2.04}). Thus we get
\begin{equation}
\label{2.06}
  \omega_{max} = \sqrt{ 1 +\beta_L +4 |\lambda| },~~~ 
  \omega_{min} = \sqrt{ 1 +\beta_L -4 |\lambda| } ,
\end{equation}
that give an approximate bandwidth $\Delta \Omega \simeq 4 |\lambda| / \Omega_{SQ}$,
where 
\begin{equation}
\label{2.06a}
  \Omega_{SQ} =\frac{\omega_{SQ}}{\omega_0} =\sqrt{1 +\beta_L} .
\end{equation}

The dynamic equations (\ref{2.01}) are integrated with a fourth-order Runge-Kutta
algorithm with fixed time-stepping, and the total energy of the SQUID metamaterial,
in units of the Josephson energy $E_J$, is calculated from the expression
\begin{eqnarray}
\label{3.01}
   E_{tot} =\sum_{n,m} \left\{ \frac{\pi}{\beta} 
       \left[ \dot{\phi}_{n,m}^2 +( \phi_{n,m} -\phi_{ext} )^2 \right]  \right.
\nonumber \\  
           +[1 -\cos(2\pi \phi_{n,m})] 
\nonumber \\  
      -\frac{\pi}{\beta}  \left[ 
   \lambda_x ( \phi_{n,m} -\phi_{ext} )( \phi_{n-1,m} -\phi_{ext} ) \right.
\nonumber \\  
  +\lambda_x ( \phi_{n+1,m} -\phi_{ext} )( \phi_{n,m} -\phi_{ext} ) 
\nonumber \\  
  +\lambda_y ( \phi_{n,m-1} -\phi_{ext} )( \phi_{n,m} -\phi_{ext} )
\nonumber \\  \left. \left.  
  +\lambda_y ( \phi_{n,m} -\phi_{ext} )( \phi_{n,m+1} -\phi_{ext} ) \right] \right\}. 
\end{eqnarray}
The averaged energy in one driving period $T=2\pi /\omega$ of evolution then reads 
\begin{equation}
\label{2.07}
   <E_{tot}>_T =\frac{1}{T} \int_0^T d\tau E_{tot} (\tau) ,
\end{equation}
where $\omega =\omega_0 \Omega$ is the frequency of the applied ac flux.

\section{dc flux tuneability}
At frequencies within a narrow band, the metamaterial absorbs a large amount of energy.
For dc flux bias equal to integer multiples of $\Phi_0$ (including zero), that band 
coincides with the linear flux-wave band given in equation (\ref{2.04}).
However, for any other value of the dc flux, that band shifts downwards, down to a minimum 
attained for dc flux bias equal to odd semi-integer multiples of $\Phi_0$.
This procedure results in a very clear pattern of periodic shifting of the linear band
with the dc flux bias, which has been observed in experiments 
\cite{Butz2013a,Butz2013b,Trepanier2013}. Here the basic features of the dc flux tuneability
patterns are reproduced by integrating equations (\ref{2.01}) and using the calculated 
fluxes $\phi_{n,m}$ and voltages $\dot{\phi}_{n,m}$ to obtain the averaged energy per 
period of the ac field $<E_{tot}>_T = <E_{tot}>_T (\Omega=2\pi/T,\phi_{dc})$  from 
equations (\ref{3.01}) and (\ref{2.07}) as a function of the frequency of the ac field 
$\Omega$ and the dc flux $\phi_{dc}$. The quantity $<E_{tot}>_T (\Omega=,\phi_{dc})$
is then presented in density plots on the $\Omega - \phi_{dc}$ plane in figure 2,
fo several combinations of ac flux amplitudes $\phi_{ac}$ and coupling coefficients 
$\lambda$. In these plots, the regions of resonant absorption correspond to maxima in the 
averaged energy, which appear as minima in the corresponding plots obtained from experiments,
where the measured quantity is the frequency-dependent complex transmission amplitude
$|S_{21} (\Omega)|$. In figure 2, while the dc flux is measured in units of $\Phi_0$,
the frequencies has been transformed in natural units ($GHz$); the single-SQUID resonance
frequency is chosen to be $f_{SQ}=15~GHz$. The frequency $f_{SQ}$ can be either calculated
from the SQUID parameters or measured experimentally. The chosen value for $f_{SQ}$ is 
slightly higher than the one measured in reference \cite{Jung2013b} for similar SQUID
parameters; no attempt for an accurate fitting of the experimental data was made in this 
work.
In figure 2, the ac amplitude $\phi_{ac}$ increases from left to right, while the coupling 
$\lambda$ increases from top to bottom. Apparently, the resonant energy absorption, 
indicated by the dark regions, become 
stronger as one moves from left to right (increasing $\phi_{ac}$), since the nonlinear 
effects become more and more important.
When going from top to bottom panels, a smearing of the resonance is observed with 
increasing $|\lambda|$, along with the appearence of secondary resonances.
The latter manifest themselves as thin dark curves that are located close to the main 
shifting 
pattern, and they are better seen around half-intefger values of the applied dc flux.
The thickness of the low-transmission (dark) region in the close-to-linear regime, 
roughly corresponds to the width of the linear band, i.e., 
$\Delta \Omega \simeq 4 |\lambda| / \Omega_{SQ}$.
\begin{figure*}[ht]
\includegraphics[angle=0, width=0.9\linewidth]{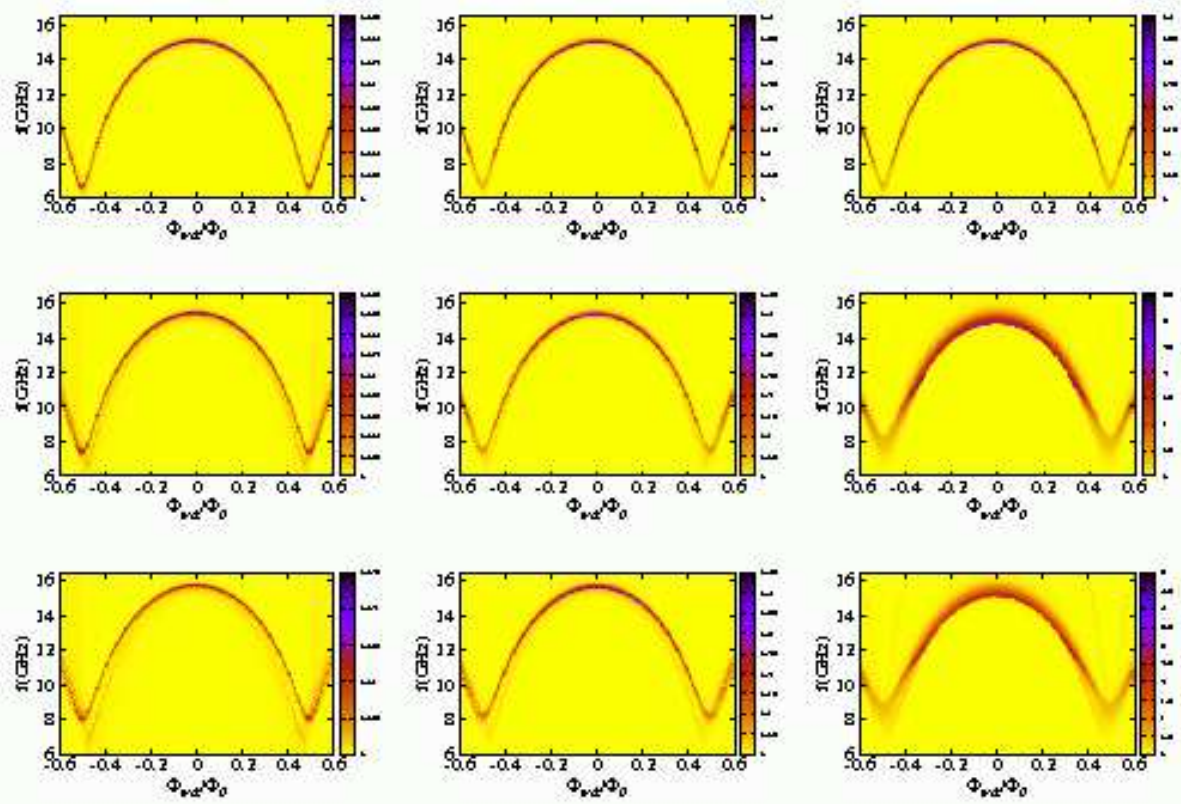}
\caption{(color online)
Density plot of the averaged total energy, $<E_{tot}>_T$
as a function of the dc flux bias $\phi_{dc}=\Phi_{dc}/\Phi_0$ and the ac frequency
$f$, for $N_x=N_y=11$, $\beta_L \simeq 0.7$, $\gamma =0.009$, and several combinations 
of ac amplitude $\phi_{ac}=\Phi_{ac}/\Phi_0$ and coupling coefficients $\lambda$.
The driving ac amplitude increases from left to right, while the coupling increases from top
to bottom.
First column:
$\lambda_x =\lambda_y =-0.01$, and $\phi_{ac}=1/5000$ (left); $1/1000$ (middle); $1/200$ (right).
Second column:
$\lambda_x =\lambda_y =-0.03$, and $\phi_{ac}=1/5000$ (left); $1/1000$ (middle); $1/200$ (right).
Third column:
$\lambda_x =\lambda_y =-0.05$, and $\phi_{ac}=1/5000$ (left); $1/1000$ (middle); $1/200$ (right).
The single-SQUID resonance frequency $f_{SQ}$ used in the calculations is $15~GHz$. 
}
\end{figure*}

These tuneability patterns can be understood within an approximate treatment valid for 
$\phi_{ac} \ll 1$. First assume that $\phi_{n,m}\simeq \phi$ for any $n,m$, i.e., that the
SQUIDs are synchronized \cite{Lazarides2013b}; small deviations from complete synchronization
arise due to the finite size of the metamaterial.
We also assume that $\gamma \simeq 0$. Substitution into equations (\ref{2.01}) 
and using equation (\ref{2.02a}), we get
\begin{equation}
\label{2.08}
   \ddot{\phi} +(1-4\lambda) \phi +\beta\, \sin(2\pi \phi) 
             =(1-4\lambda)[\phi_{dc} +\phi_{ac} \cos(\Omega \tau)]. 
\end{equation}
In the earlier equation we further make the approximation
\begin{equation}
\label{2.08.1}
  \beta \sin(2\pi \phi) \simeq \beta_L \phi -\frac{2\pi^2}{3} \beta_L \phi^3 ,
\end{equation}
and the ansatz 
\begin{equation}
\label{2.08.2}
\phi =\phi_0 +\phi_1  \cos(\Omega \tau ) ,
\end{equation} 
Substituting equations (\ref{2.08.1}) and (\ref{2.08.2}) into equation (\ref{2.08}),
using the rotating wave approximation (RWA), and separating constant from 
time-dependent terms, we get
\begin{eqnarray}
\label{2.09}
   \frac{2\pi^2}{3} \beta_L \phi_0^3 -(1-4\lambda+\beta_L) \phi_0 -\frac{3}{2}\phi_0 \phi_1^2
\nonumber \\
    +(1-4\lambda) \phi_{dc} =0 , \\
\label{2.10}
  \frac{\pi^2}{2} \beta_L \phi_1^3 
     -\left\{ (1-4\lambda+\beta_L-\Omega^2) -2\pi^2 \beta_L \phi_0^2 \right\} \phi_1
\nonumber \\
     +(1-4\lambda) \phi_{ac} =0 .
\end{eqnarray}
Limiting ourselves in the case $\phi_1 < \phi_0 <<1$, we may simplify equations (\ref{2.09})
and (\ref{2.10}), by neglecting terms proportional to $\phi_1^3$, $\phi_0^3$, and
$\phi_0 \phi_1^1$. Note that we keep the term $\propto \phi_0^2 \phi_1$, i.e., the lowest
order coupling term between the two equations. Then, the resulting equations can be easily
solved to give
\begin{eqnarray}
\label{2.11}
   \phi_0 = \frac{(1-4\lambda) \phi_{dc}}{(\Omega_{SQ}^2 -4\lambda)} ; \\ 
\label{2.11a}
   \phi_1 = \frac{(1-4\lambda) \phi_{ac}}{\left\{ (\Omega_{SQ}^2 -4\lambda -\Omega^2) -2\pi^2 \beta_L \phi_0^2 \right\}} .
\end{eqnarray}
Obviously, the ac flux amplitude in the SQUIDs, $\phi_1$, attains its maximum value when
the expression in the curly brackets in the denominator of equation (\ref{2.11a}) is zero.
Solving for $\Omega$, we get
\begin{equation}
\label{2.12}
   \Omega =\sqrt{ (\Omega_{SQ}^2 -4\lambda) -(2\pi^2 \beta_L)
       \frac{(1-4\lambda)^2 \phi_{dc}^2}{(\Omega_{SQ}^2 -4\lambda)^2} } ,
\end{equation}
or, in natural units
\begin{equation}
\label{2.13}
   f =\frac{f_{SQ}}{\Omega_{SQ}} \sqrt{ (\Omega_{SQ}^2 -4\lambda) -(2\pi^2 \beta_L)
       \frac{(1-4\lambda)^2 \phi_{dc}^2}{(\Omega_{SQ}^2 -4\lambda)^2} } ,
\end{equation}
which corresponds to the "resonance frequency" of the SQUID metamaterial itself,
with $f_{SQ}$ is the single-SQUID resonance frequency.
\begin{figure}[!t]
\includegraphics[angle=0, width=0.80\linewidth]{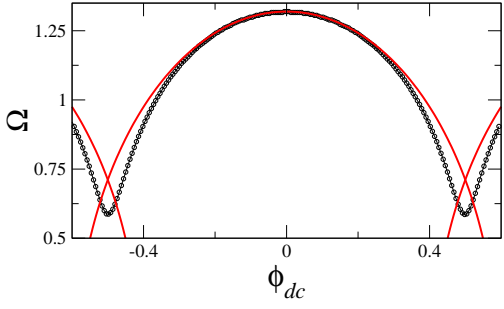}
\caption{(color online)
Normalized frequency at maximum response of the SQUID metamaterial, $\Omega$, as a function
of the dc applied (normalized) flux, $\phi_{dc}$, in the presence of a low-amplitude
alternating signal. The black circles are obtained from the numerical simulations through
model equations (\ref{2.01}) (see text), while the red solid lines are plotted from the 
approximate equation (\ref{2.12}).
Parameters:
$\lambda_x =\lambda_y =-0.01$, $N_x=N_y=11$, $\phi_{ac}=1/5000$, $\gamma =0.009$, and
$\beta_L \simeq 0.7$.
}
\end{figure}
From the data of the tuneability patterns presented in the left column of figure 2
(weak driving amplitude), that correspond to increasing coulping from top to bottom,
we have extracted the frequency of maximum response by simply identifying the frequency
where the total energy is maximum. A typical curve of this kind is shown in figure 3,
along with the corresponding one calculated from equation (\ref{2.12}).
This is the case of low-amplitude ac driving, that is the closest to the assumptions
made for the derivation of expression (\ref{2.12}). It is remarkable that this simple
expression, which contains only two parameters, $\lambda$ and $\beta_L$, fairly agrees
with the simulations in a rather wide region of dc fluxes,
i.e., from $\phi_{dc} \sim -0.3$ to $\sim +0.3$. Within this interval,
the normalized frequency in figure 3 changes from $\Omega=1.12$ to $1.32$,
that makes a tuneability range of $~15\%$. This is approximatelly the "useful" tuneability
range \cite{Butz2013a,Butz2013b}, since at these frequencies the energy absorption remains 
at high levels (i.e., the resonance is strong).
For larger $\phi_{dc}$, the importance of the term $\propto \phi_0^3$ increases and it
cannot be neglected for the solutions of equations (\ref{2.09}) and (\ref{2.10}).
The approximate expression (\ref{2.12}) also captures another experimentally observed
feature, namely the increase of the metamaterial resonance frequency at zero dc flux,
$f (\phi_{dc}=0)$. By setting $\phi_{dc} =0$ in equation (\ref{2.12}) we get that 
$f =\frac{f_{SQ}}{\Omega_{SQ}} \sqrt{ (\Omega_{SQ}^2 -4\lambda) }$, which, for 
$f_{SQ} =15~GHz$, $\Omega_{SQ} =1.304$ ($\beta_L=0.7$), $\lambda =-0.01, ~-0.03, ~-0.05$
gives respectively, $f=15.2, ~15.5, 15.9 ~GHz$ in agreement with the numerical results. 
The $\lambda-$dependence of the SQUID metamaterial resonance frequency
is weaker in the corresponding one-dimensional case.  
This effect does not however result directly from the nonlinearity; the ac field is very
week to induce significant nonlinear effects. Instead, it comes from the assumed uniformity
of the SQUID metamaterial state, i.e., the assumption $\phi_{n,m} =\phi$ for any $n,m$.
From that, and neglecting the dissipation and driving terms, we get the single 
eigenfrequency of the metamaterial in that state as $\Omega =\sqrt{\Omega_{SQ} -4\lambda}$,
so that deviations of the resonance frequency from that of a single SQUID are 
approximatelly proportional to $\lambda$ ($|\lambda| \ll 1$).

\section{Energy Transmission through SQUID Metamaterial Lines.-}
SQUID metamaterials support flux waves that are capable of transmitting energy, 
in much the same way as in nonlinear magnetoinductive transmission lines made of
conventional (metallic) metamaterials \cite{Lazarides2011a}, which may function as a
frequency-selective communication channel for devices.
For simplicity and clarity we use a one-dimensional SQUID metamaterial comprising 
$N=54$ identical elements with $\beta_L =0.7$ ($\beta=0.1114$) weakly coupled to their 
nearest neighbors. In order to investigate the transfer of energy through the array,
the SQUID that is located at the left end (i.e., that for $n=1$) is excited by an ac flux 
field at a particular frequency.
The system of equations (\ref{2.01}) and equation (\ref{3.01}) in the present case read
\begin{eqnarray}
\label{4.01}
  \ddot{\phi}_{n} +\gamma \dot{\phi}_{n} +\phi_{n} +\beta\, \sin( 2 \pi \phi_{n} )
  -\lambda ( \phi_{n-1} +\phi_{n+1} ) 
\nonumber \\
   = (1-2\lambda) \phi_{ext} \, \delta_{n,1} ,
\end{eqnarray}
where the coupling coefficient is now denoted by $\lambda$ and the Kroneckers' delta
$\delta_{n,1}$ indicates that only the SQUID with $n=1$ is driven by the ac field,
and 
\begin{eqnarray}
\label{4.02}
   E_{tot} =\sum_{n} \left\{ \frac{\pi}{\beta} 
       \left[ \dot{\phi}_{n}^2 +( \phi_{n} -\phi_{ext} )^2 \right]  \right.
           +1-\cos(2\pi \phi_{n}) 
\nonumber \\  
  -\frac{\pi}{\beta}  \left[ 
   \lambda ( \phi_{n} -\phi_{ext} )( \phi_{n-1} -\phi_{ext} ) \right. 
\nonumber \\  \left. \left.
  +\lambda ( \phi_{n+1} -\phi_{ext} )( \phi_{n} -\phi_{ext} ) \right] \right\}  
\end{eqnarray}
Equations (\ref{4.01}) implemented with the boundary conditions $\phi_0 =\phi_{N+1}$
are integrated in time for 
$12000\, T$ time units, where $T=2\pi/\Omega$ is the driving period, so that transient 
effects are eliminated and the system has reached a stationary state.
The energy density in the metamaterial is then calculated from (\ref{4.02}) as the 
average over the next
$2000\, T$ time units. The decimal logarithm of the averaged energy density is mapped on 
the frequency $\Omega$ - site number $n$ plane (figure 4), where high transmission regions
are indicated with darker colors. In figure 4, the changes in the energy transmission with 
respect to the dissipation coefficient $\gamma$ are shown for fixed coupling coefficient 
$\lambda=-0.01$ and ac field $\phi{ac}=0.1$. Note that for that value of $\phi{ac}$,
significant nonlinearity is already present.

The energy transmission map for relatively strong dissipation ($\gamma=0.009$) is shown 
in the upper panel of figure 4. Significant energy transmission occurs in a narrow 
band, of the order $\sim 2\lambda$ around the single SQUID resonance frequency
$\Omega_{SQ}\simeq 1.3$ (for $\beta_L=0.7$). This band almost coincides with the linear
band for the one-dimensional SQUID metamaterial. Note that energy transmission also
occurs at other frequencies; e.g., at $\Omega \sim 0.43$ that corresponds to a 
subharmonic resonance. Subharmonic resonances result from nonlinearity; in this case,
nonlinear effects are already significant due to the relatively high ac field amplitude
($\phi_{ac} =0.1$). 
However, with decreasing losses (middle panel), more energy is transmitted both at
frequencies in the linear band and the subharmonic resonance band. In the following we 
refer to the latter as the nonlinear band, since it results from purely nonlinear effects.
With further decrease of losses (lower panel), the transmitted energy in these two bands
becomes more significant. The comparison can be made more clearly by looking at the panels
in the middle and right columns, which show enlarged regions of the corresponding panels 
shown in the left column. The enlargement around the linear band shown in the middle 
column shows clearly the increase of the transmitted energy with decreasing losses.
Moreover, in the case of very low losses ($\gamma=0.001$) the linear band splits into two
bands, where significant energy transmission occurs.
The energy transmission in the nonlinear band increases with decreasing losses, 
accordingly (left column, losses decrease from top to bottom). 
\begin{figure*}[!t]
\includegraphics[angle=0, width=0.70\linewidth] {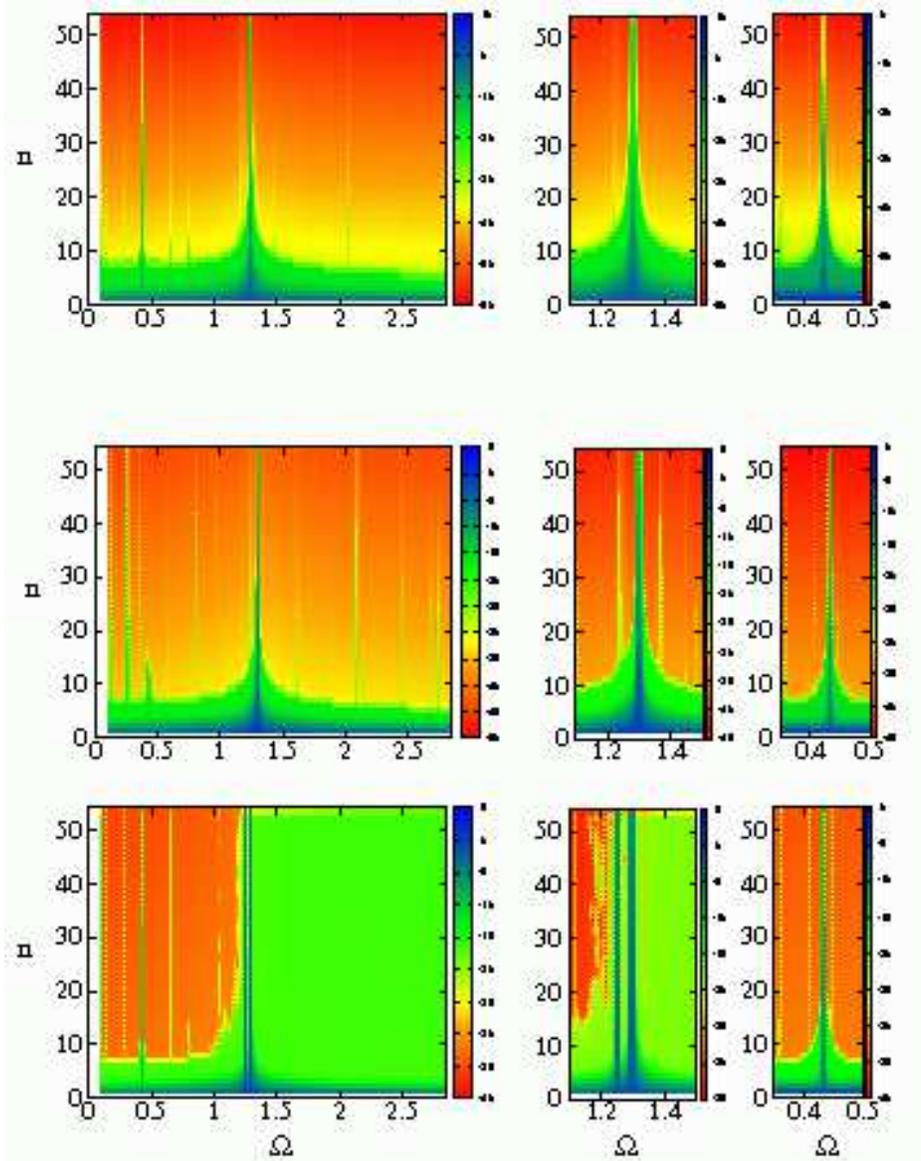}
\caption{(color online)
Energy transmission through a one-dimensional SQUID metamaterial line with $N=54$ SQUIDs.
The logarithm of the energy density averaged over $2000\, T$ time units,
$\log_{10} [ <E_n>_T ]$, is plotted on the site number $n$ - driving frequency $\Omega$
plane for $\beta_L =0.7$ ($\beta=0.1114$), $\lambda=-0.01$, and $\gamma=0.009$ (upper);
$\gamma=0.004$ (middle);  $\gamma=0.001$ (lower). The middle and left columns are
enlargements in frequency ranges around the fundamental and the subharmonic resonance,
respectively, at $\Omega \simeq 1.302$ and $0.43$.  
}
\end{figure*}

Thus, for sufficiently strong ac field, energy can be transmitted not only for 
frequencies in the linear band, but also in otherwise forbidden frequency regions.
The metamaterial thus becomes transparent in energy transmission at frequency intervals 
around nonlinear resonances (subharmonic, in the present case) of single rf SQUIDs.
That type of self-induced transparency due to nonlinearity is a robust effect as can be 
seen in figure 4,
where the loss coefficient has been varied by almost an order of magnitude. 
In the density plots above, the right boundary was actually a reflecting one,
that allows the formation of stationary states in the array. However, the same calculations
performed with a totally absorbing boundary give practically identical results.

For very low amplitude of the applied ac field $\phi_{ac}$ and $\phi_{dc}=0$, 
equations (\ref{4.01}) can be linearized to give 
\begin{eqnarray}
\label{4.03}
  \ddot{\phi}_{n} +\gamma \dot{\phi}_{n} +\Omega_{SQ}^2 \phi_{n}
  -\lambda ( \phi_{n-1} +\phi_{n+1} ) \nonumber \\
  =\bar{\phi}_{ac} \cos(\Omega \tau) \, \delta_{n,1} ,
\end{eqnarray}
where $\bar{\phi}_{ac} =(1-2\lambda) \phi_{ac}$. If we further neglect the loss term,
equations (\ref{4.03}) can be solved exactly in closed form for any driving frequency 
$\Omega$ and for any finite $N$, the total
number of SQUIDs in the one-dimensional array. By substitution of the trial solution 
$\phi_n = q_n \cos(\Omega \tau)$ into equations (\ref{4.03}) and after some rearrangement 
we get 
\begin{equation}
\label{4.04}
  s q_{n-1} + q_n + s q_{n+1} = \kappa_0 \, \delta_{n,1} ,
\end{equation}
where
\begin{equation}
\label{4.05}
  s=-\frac{\lambda}{\Omega_{SQ}^2 -\Omega^2}, \qquad
  \kappa_0 =\frac{\bar{\phi}_{ac}}{\Omega_{SQ}^2 -\Omega^2} ,
\end{equation}
or, in matrix form
\begin{equation}
  \label{4.06}
    {\bf q} = \kappa_0 \, \hat{\bf S}^{-1} {\bf E}_1 ,
\end{equation}
where  ${\bf q}$ and ${\bf E}_1$ are $N-$dimensional vectors  with componets $q_n$
and $\delta_{n,1}$, respectively, and $\hat{\bf S}^{-1}$ is the inverse of the 
$N\times N$ coupling matrix $\hat{\bf S}$. The latter is a real, symmetric, tridiagonal 
matrix that has its diagonal elements equal to unity, while all the other non-zero 
elements are equal to $s$. The elements of the matrix $\hat{\bf S}^{-1}$ can be 
obtained in closed analytical form \cite{Lazarides2010a} using known results for the 
inversion of more general tridiagonal matrices \cite{Huang1997}. 
Then, the components of the ${\bf q}$ vector can be written as
\begin{equation}
  \label{4.07}
   q_n = \kappa_0 \left( \hat{\bf S}^{-1} \right)_{n,1} ,
\end{equation}
where $\left( \hat{\bf S}^{-1} \right)_{n,1}$ is the $(n,1)-$element of 
$\hat{\bf S}^{-1}$, whose explicit form is given in reference \cite{Lazarides2010a}.
Then, the solution of the linear system (\ref{4.03}) with $\gamma=0$ is  
\begin{eqnarray}
  \label{4.08} 
    \phi_n (\tau) = \kappa_0 \mu \frac{\sin[(N-n+1)\theta']}{\sin[(N+1)\theta']}
       \cos(\Omega \tau) ,
\nonumber \\
  \theta' = \cos^{-1} \left( \frac{1}{2|s|} \right) , 
\end{eqnarray}
for $s>+1/2$ and $s<-1/2$ (in the linear flux-wave band),
and
\begin{eqnarray}
  \label{4.09} 
    \phi_n (\tau) = \kappa_ \mu \frac{\sinh[(N-n+1)\theta]}{\sinh[(N+1)\theta]} 
    \cos(\Omega \tau) ,
\nonumber \\
  \theta = \ln\frac{1+\sqrt{1-(2s)^2}}{2|s|} , 
\end{eqnarray}
for $-1/2 < s < +1/2$ (outside the linear flux-wave band),
where
\begin{eqnarray}
  \label{4.10}
   \mu= \frac{1}{|s|} \left( -\frac{|s|}{s} \right)^{n-1} .
\end{eqnarray}
The above expressions actually provide the asymptotic solutions, i.e., after the transients
due to dissipation, etc., have died out. Thus, these driven linear modes correspond to the 
stationary state of the linearized system; the dissipation however may alter somewhat their
amplitude, without affecting very much their form.
Note also that the $q_n$s are uniquely determined by the parameters of the system, and they
vanish with vanishing $\phi_{ac}$.

From the analytical solution at frequencies within the linear flux-wave band, equation
(\ref{4.08}) which corresponds to either $s>+1/2$ or $s<-1/2$,
the resonance frequencies of the array can be obtained by setting $\sin[(N+1)\theta'] =0$.
Thus we get  
\begin{eqnarray}
  \label{4.11}
  s\equiv s_m=\frac{1}{2 \cos\left[ \frac{m\pi}{(N+1)} \right]} ,
\end{eqnarray}
where $m$ is an integer ($m=1,...,N$). By solving the first of equations (\ref{4.05}) 
with respect to $\Omega$, and substituting the values of $s\equiv s_m$ from equation 
(\ref{4.11}) we get
\begin{eqnarray}
  \label{}
   \Omega \equiv \Omega_m =
       \sqrt{ \Omega_{SQ}^2 +2\, \lambda \, \cos\left( \frac{m\pi}{N+1} \right) } ,
\end{eqnarray}
which is the discrete frequency dispersion for linear flux-waves in a one-dimensional SQUID
metamaterial, with $m$ being the mode number ($m=1,...,N$).

\section{Dynamic Multistability and Power-Dependent Tuneability.-}
The dc flux tuneability of SQUID metamaterials is a truly nonlinear effect, which appears
also in the "linear regime", where the very low ac power levels allow for the treatment 
of a Josephson junction as a quasi-linear inductance. In figure 2 we have observed that
with increasing $\phi_{ac}$ (increasing thus the significance of nonlinearity), the 
complexity of the tuneability patterns increases as well, while they exhibit high power
absorption in a frequency range that exceeds the boundaries of the linear band.
Thus, with increasing power, new dynamic effects are expected to appear;  
with resonance frequency shifts and multistability being the most prominent.
The latter, in particular, is a purely dynamic phenomenon that is not related to the 
multistability known from hysteretic SQUIDs. For a particular choise of parameters,
dynamic multistability manifests itself as a small number of simultaneously stable states.
As it has been shown in recent experiments \cite{Jung2013}, each of these states corresponds
to a different value of the SQUID magnetic flux susceptibility at the driving frequency,
that may be either possitive or negative.
This implies that, depending on the state of individual SQUIDs, the metamaterial can either 
be magnetically almost transparent or not.

While in figure 2 the amplitude of the ac field was kept to low values ($0.005$ the highest),
in the present section we use higher values. Due to the resonant nature of the SQUIDs, the 
current may attain values that are higher than the critival current of the Josephson 
junctions. In this case, the standard RCSJ model gives unphysically high current values, 
because it does not take into account the change in the conductance of the junctions
\cite{Likharev1986} (p. 48). In order to incorporate that change, we replace the dissipation
coefficient $\gamma$ in equations (\ref{2.01}) with the following current dependent function
\begin{equation}
  \label{5.01}
   \gamma_{n,m}=\gamma_0 +c \left[ 1 +\tanh \left(\frac{i_{n,m}-1}{d} \right) \right] ,
\end{equation}
where $c=0.12$, $d=0.02$, and $\gamma_0 =0.009$, the value of $\gamma$ used for obtaining 
the results of figure 2. That function allows for an abrupt but continuous change of the 
dissipation 
coefficient from low values $\gamma_{n,m} \simeq \gamma_0$ (low current, less than $I_c$)
to high values $\gamma_{n,m} \simeq 28 \gamma_0$ (high current, greater than $I_c$).
We have calculated several curves of the total current maximum $i_{max}$ as a function
of the driving frequency $\Omega$, which are shown in figure 5.
In this figure, the amplitude of the applied ac field increases from the bottom  to the 
upper panel [from (f) to (a)]. In all subfigures, the frequency is varied 
between $\Omega \simeq 0.8 -6.28$ in both increasing from the lowest frequency or decreasing
from the highest frequency (only part of this range is shown). In this way we may identify 
two different branches of the $i_{max} - \Omega$ at a particular frequency range for relatively
high ac powers.
For low power of the ac field [figures 5(f)-5d] no hysteresis and thus multistability
is observed. However, hysteretic effects appears in figure 5(c) for $\phi_{ac} =0.03$ 
for a very narrow frequency interval. With further increasing $\phi_{ac}$ we see that
the hysteretic lobe, and thus the frequency region of multistability, significantly 
increases [figures 5(b)-5(a)]. 
Besides the multistability effects, in figure 5 we also observe strong resonance red-shift
with increasing ac power level. For low power levels [figure 5(f)] the curve exhibits a 
strong resonant response at $\sim \Omega_{SQ}$, the single SQUID (linear) resonance frequency.
With increasing ac power, however, the resonance frequency, determined by the maximum of
each curve, moves to lower frequencies. Note that in figures 5(b) and 5(a), there are 
frequency regions where the SQUID metamaterial jumps intermittently from the low to the 
high current state (or vice versa). In comparison with the standard RCSJ model,
that nonlinear resistive Josephson junction model severely limits total maximum currents 
which are greater 
than the critical one; it also exhibit much less hysteretic effects, that are visible 
at relatively high powers only.
\begin{figure}[!t]
\centerline{\epsfig{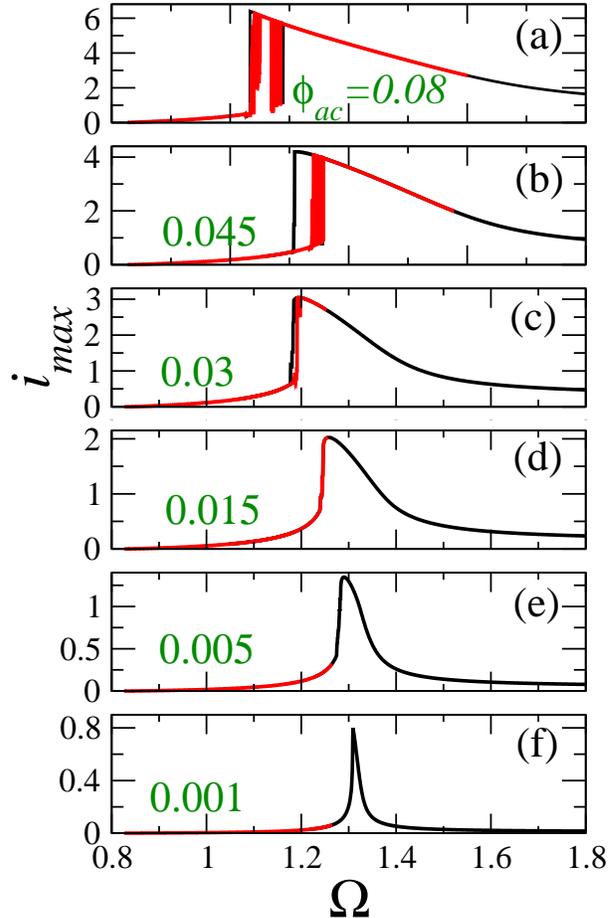}}
\caption{(color online)
The maximum total current $i_{max}$ in the SQUID metamaterial as a function of the normalized
frequency $\Omega$, for $N_x=N_y=11$, $\lambda_x =\lambda_y =-0.01$, $\beta_L \simeq 0.7$,
$\gamma_0 =0.009$ (see text), $\phi{dc}=0$, and (a) $\phi_{ac}=0.08$; (b) $\phi_{ac}=0.045$;
(c) $\phi_{ac}=0.03$; (d) $\phi_{ac}=0.015$; (e) $\phi_{ac}=0.005$; (f) $\phi_{ac}=0.001$.
}
\label{squid-rf-tuneability}
\end{figure}

\section{Conclusions.-}
We have used a SQUID metamaterial model in the weak coupling approximation, in order to 
explore the tuneability, nonlinear transmission, and dynamic multistability properties of
SQUID metamaterials.
The numerical results reproduce very well the experimentally observed dc flux tuneability 
patterns, whose form remains qualitatively unaffected from the details of the particular 
SQUID metamaterial. System parameters such as the size of the system, its dimensionality,
the SQUID parameter, and the coupling between SQUIDs, mainly affect the tuenability range
(i.e., the minimum and maximum value of resonance frequency of the metamaterial).

In the linear regime, a simple approximate expression which gives the dependence of the 
resonance frequency on the dc flux bias and the coupling strength can be obtained.
Although it has been obtained under several simplifying assumptions, it agrees fairly well 
with the numerically obtained tuneability patterns within a significant tuneability range.
This expression also explains the increase of the zero dc flux resonance frequency with 
increasing coupling strength, which comes from the excitation of a mode where all the 
SQUIDs are synchronized. 

Simulations of one-dimensional SQUID metamaterials which are driven at one end, reveal 
their energy transmission properties. For low amplitude of the ac applied field, 
energy transmission is limited at frequencies within the linear band. In that case,
the linearized equations can be solved exactly, and the amount of transmitted energy in 
each SQUID could be calculated (with a correction for slight reduction of energy due to 
dissipation which has been neglected in the analytical calculations).
For stronger ac fields, the nonlinear effects become significant, and significant amounts
of energy can be transmitted through nonlinear frequency bands which appear due to 
secondary SQUID resonances. 
 
The resonance of the maximum total current as a function of the driving frequency shifts
to lower frequencies with increasing ac field amplitude. For relatively strong ac fields,
dynamic bistability appears. For a more realistic description of the Josephson junctions
in the SQUID metamaterial, the nonlinear variation of their resistance with current in
each SQUID has been taken into account.

\begin{acknowledgement}
This work was partially supported by
the European Union's Seventh Framework Programme (FP7-REGPOT-2012-2013-1)
under grant agreement n$^o$ 316165, and
by the Thales Project MACOMSYS, co‐financed by the European Union
(European Social Fund – ESF) and Greek national funds through the Operational
Program "Education and Lifelong Learning" of the National Strategic Reference
Framework (NSRF) ‐ Research Funding Program: THALES.
Investing in knowledge society through the European Social Fund.
\end{acknowledgement}


\end{document}